\documentclass[sigconf]{acmart}

\usepackage{soul}
\usepackage{fontawesome5}
\usepackage{algorithm2e}
\usepackage{hyperref}
\usepackage{balance}
\usepackage{makecell}
\usepackage{siunitx}

\newcommand{\cmt}[1]{\ignorespaces}

\SetKwInput{KwInput}{Input}
\SetKwInput{KwOutput}{Output}
\SetKwInput{KwName}{Name}

\usepackage{mathtools}

\usepackage{multirow}

\AtBeginDocument{%
  }

\copyrightyear{2026}
\acmYear{2026}
\setcopyright{cc}
\setcctype{by}
\acmConference[CHI EA '26]{Extended Abstracts of the 2026 CHI Conference on Human Factors in Computing Systems}{April 13--17, 2026}{Barcelona, Spain}
\acmBooktitle{Extended Abstracts of the 2026 CHI Conference on Human Factors in Computing Systems (CHI EA '26), April 13--17, 2026, Barcelona, Spain}
\acmDOI{10.1145/3772363.3798434}
\acmISBN{979-8-4007-2281-3/2026/04}

\begin{document}

\title[Sketch2Topo]{Sketch2Topo: Using Hand-Drawn Inputs for Diffusion-Based Topology Optimization}


\author{Shuyue Feng}
\orcid{0000-0002-6720-5356}
\affiliation{%
  \institution{The University of Tokyo}
  \city{Tokyo}
  \country{Japan}
}
\email{shuyuefeng@akg.t.u-tokyo.ac.jp}

\author{Cedric Caremel}
\orcid{0000-0002-3547-7285}
\affiliation{%
  \institution{The University of Tokyo}
  \city{Tokyo}
  \country{Japan}
}
\email{cedric@akg.t.u-tokyo.ac.jp}

\author{Yoshihiro Kawahara}
\authornote{Corresponding author.}
\orcid{0000-0002-3547-7285}
\affiliation{%
  \institution{The University of Tokyo}
  \city{Tokyo}
  \country{Japan}
}
\email{kawahara@akg.t.u-tokyo.ac.jp}

\renewcommand{\shortauthors}{Feng et al.}

\begin{abstract}
Topology optimization (TO) is employed in engineering to optimize structural performance while maximizing material efficiency. However, traditional TO methods incur significant computational and time costs. Although research has leveraged generative AI to predict TO outcomes and validated feasibility and accuracy, existing approaches still suffer from limited customizability and impose a high cognitive load on users. Furthermore, balancing structural performance with aesthetic attributes remains a persistent challenge. We developed Sketch2Topo, which augments a diffusion-based TO model with image-to-image generation and image editing capabilities. With Sketch2Topo, users can use sketching to customize geometries and specify physical constraints. The tool also supports mask input, enabling users to perform TO on selected regions only, thereby supporting higher levels of customization. We summarize the workflow and details of the tool and conduct a brief quantitative evaluation. Finally, we explore application scenarios and discuss how hand-drawn input improves usability while balancing functionality and aesthetics.
\end{abstract}



 \begin{teaserfigure}
   \includegraphics[width=\textwidth]{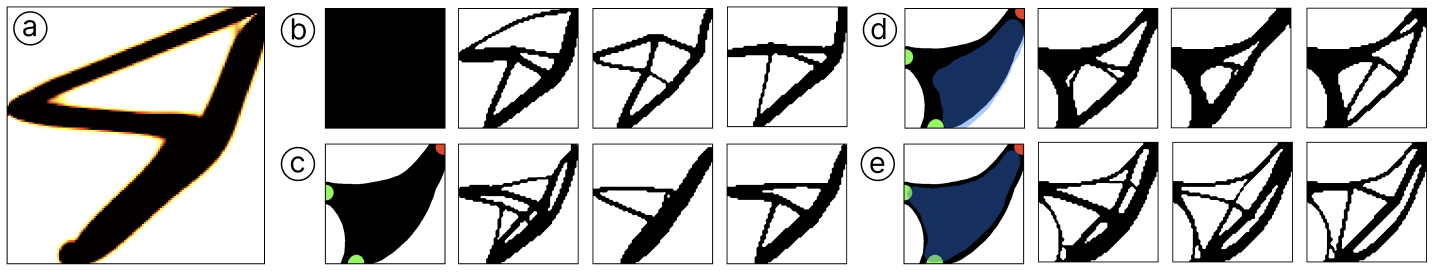}
   \caption{Results obtained under identical constraints using different topology optimization approaches. (a) Traditional finite element analysis based on the SIMP algorithm. (b) Prior generative-AI-based topology optimization. (c) Image-to-image generation provided by Sketch2Topo. (d) and (e) Inpainting-based generation provided by Sketch2Topo.}
   \Description{In Figure 1.}
  \label{Figure1}
 \end{teaserfigure}

\maketitle

\section{Introduction}

Topology Optimization (TO) is a fundamental technique in modern engineering and industrial design, aiming to maximize structural performance with minimal material usage~\cite{bendsoe2003topology}. However, traditional topology optimization workflows still face significant challenges in the early stages of design. This process is typically accompanied by extremely high computational and temporal costs, often requiring hours to converge to a single solution~\cite{aage2017giga}. The high computational and time costs make it difficult for designers to rapidly iterate and compare multiple alternatives while also considering aesthetics.

On the other hand, some researchers have turned to utilizing generative AI (particularly diffusion models) to predict topology optimization results~\cite{maze2023diffusion,giannone2023aligning,nobari2024nito}. Although these data-driven methods have successfully validated the effectiveness of generating high-precision and high-speed TO structures, they have severe limitations in usability. First, existing methods treat the entire image (usually a square) as material and perform topology optimization prediction~\cite{maze2023diffusion}. This approach lacks the flexibility to handle specific geometric shapes, as the actual process often requires optimizing components of specific shapes. Second, the interaction mechanism remains non-intuitive; physical constraints (such as loads and boundary conditions) are usually input via abstract parameters or text embeddings, which are separated from the visual representation. This disconnect creates a ``black-box'' experience, which leads to a high cognitive load during use.~\cite{chen2018forte}.

To bridge these gaps, we propose Sketch2Topo, an interactive design tool that integrates generative AI into a sketch interaction workflow. By extending image-to-image and inpainting functionalities on a foundation diffusion model, we developed a system that restores the ``human-in-the-loop''~\cite{choi2021ilvr,saharia2022palette}. Our tool provides a direct manipulation interface where users can use brushes to draw component shapes and intuitively draw physical constraints on the canvas. Furthermore, Sketch2Topo incorporates an optional Mask tool to designate specific regions for local topology optimization. This allows for precise customization without disrupting the surrounding structure.

We present the implementation details of the tool and conduct a brief quantitative evaluation of minimum compliance and volume fraction. In addition, we explore potential application scenarios to demonstrate its potential in improving usability and balancing functionality with aesthetic qualities. We discuss how sketching, as a natural interaction interface, not only accelerates the engineering workflow but also enhances the aesthetic attributes of structural design compared to traditional methods.

The main contributions of this work are as follows:
\begin{itemize}
  \item Propose an interactive design workflow, Sketch2Topo, which enables topology optimization generation by using sketches to input target objects and physical constraints, and allows for the precise selection of specific regions for topology optimization;
  \item The quantitative evaluation, together with the exploration of application scenarios, aims to demonstrate the advantages of Sketch2Topo in enhancing the usability and aesthetic attributes of topology optimization.
\end{itemize}

\section{Related Work}

\begin{figure*}
\centering
\includegraphics[width=0.8\linewidth]{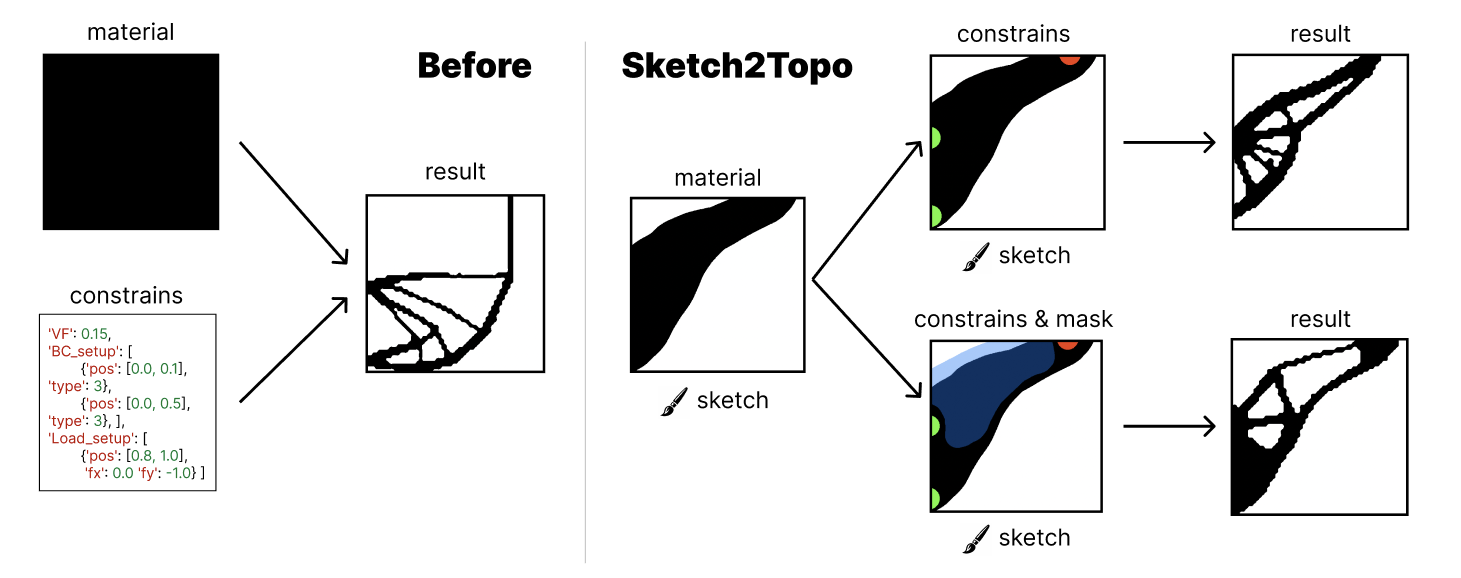}
\caption{The left panel illustrates the workflow in prior work, where conditions are specified via text-based parameters and users cannot provide a specific target shape as an input condition. The right panel shows the Sketch2Topo workflow: users can sketch the target shape and directly specify physical constraints through hand-drawn inputs. In addition, users can use a masking tool to restrict topology optimization to selected regions.}
\label{Figure2}
\end{figure*}

\subsection{Topology Optimization and Generative AI}

Topology optimization aims to optimize the material distribution within a given design domain under constraints to maximize structural performance~\cite{bendsoe2003topology,wu2015system}. Classical algorithms, such as Solid Isotropic Material with Penalization~\cite{bendsoe1989optimal} and the Level Set Method~\cite{osher2001level}, rely on iterative Finite Element Analysis (FEA) to continuously update physical fields~\cite{turner1956stiffness}. Although these methods ensure strict physical validity, they are computationally prohibitive and slow to converge~\cite{aage2017giga}. To overcome this computational bottleneck, researchers have turned to deep learning. Early works utilized Convolutional Neural Networks~\cite{yu2019deep} and Generative Adversarial Networks~\cite{nie2021topologygan} to model topology optimization as an image translation task, achieving near real-time structural prediction. Recently, diffusion models have further enhanced the diversity and high-fidelity details of generated structures~\cite{maze2023diffusion, giannone2023aligning}. However, most existing generative methods still operate in a "black box" mode, where users can only input abstract global parameters (e.g., loads, volume fraction) without the ability to intervene in the generation process~\cite{chen2018forte, matejka2018dream}. This results in a lack of flexibility, making it difficult to integrate specific geometric constraints or perform local modifications. In this work, we introduce user sketches into the generative loop, aiming to combine the intuitive control of traditional methods with the efficiency of generative models.

\subsection{Controllable Editing in Diffusion Models}

Although large-scale text-to-image diffusion models have revolutionized visual content creation, relying solely on text prompts lacks the spatial precision required for engineering design~\cite{liu2022design,chong2025prompting}. To bridge this gap, various controllable generation techniques have emerged. Methods such as ControlNet and T2I-Adapter introduce additional encoding layers to modulate the generation process based on spatial signals (e.g., Canny edges, depth maps, or user sketches)~\cite{zhang2023adding,mou2024t2i}. Furthermore, inpainting techniques allow models to regenerate specific masked regions while preserving the surrounding context, thereby enabling local editing~\cite{choi2021ilvr,saharia2022palette}. DDIM Inversion is often employed to invert images back into the initial noise space to maximize structural consistency and preserve the original features of non-edited regions during modification~\cite{mokady2023null}. Although these methods have been widely explored in artistic and media domains, their application in functional structural design remains limited. We adapt these controllable editing mechanisms to the domain of topology optimization. This enables our system to interpret freehand input not merely as visual references, but as functional definitions of physical constraints and material distribution.

\section{system overview}

\subsection{Functionality and User Interface}

Previous research on diffusion-based topology optimization typically treats the entire image (usually a square) as the material domain. However, real-world demands often involve components with specific shapes, as shown in the comparison of the material regions in \autoref{Figure2}. Therefore, we introduce an image-to-image generation approach, which first adds a certain level of noise to the input image and then performs guided generation during the denoising process~\cite{meng2021sdedit}. This enables the model to preserve and redraw the original structure without requiring any fine-tuning. As a result, users can specify target shapes using a brush-based interface, rather than relying solely on traditional parameter inputs, allowing the system to handle arbitrary shapes beyond simple square geometries. Furthermore, inputting physical constraints independently from the generation process is often non-intuitive; thus, we enable users to input physical constraints directly via sketches, as shown in the upper-right part of \autoref{Figure2}. Finally, to enhance customizability, we integrated an inpainting function, allowing users to perform topology optimization only on specific regions, as shown in the lower-right part of \autoref{Figure2}. This increases the controllability of the process, facilitating the creation of solutions that balance both performance and aesthetics.

\begin{figure*}
\centering
\includegraphics[width=0.8\linewidth]{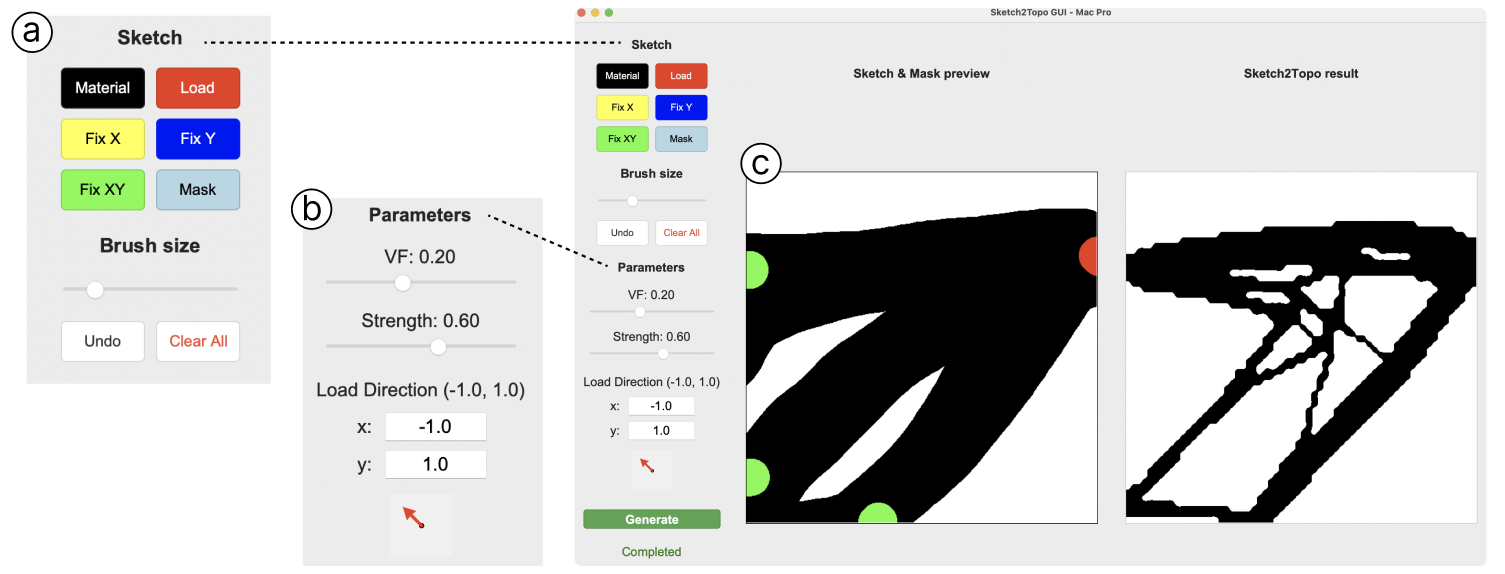}
\caption{Sketch2Topo GUI. (a) Brush tools for drawing the material region and physical constraints. (b) Parameters that are less suitable for sketch input remain text-based, allowing users to set the volume fraction and load direction. (c) Canvas area: the left canvas is used to draw the material and physical constraints, while the right canvas displays the generated result.}
\label{Figure3}
\end{figure*}

We structured the user interface as a digital painting-like environment, effectively lowering the entry barrier for users familiar with standard graphic design tools (as shown in \autoref{Figure3}). The graphical interface consists of three main parts. The first part is the sketch brush area (\autoref{Figure3}a). To translate visual inputs into engineering constraints, we designed a set of color-coded brushes. The black brush defines the initial material domain or specific geometries; the red brush specifies regions subjected to external forces; while yellow, blue, and green brushes represent fixed boundary conditions in the x-direction, y-direction, and both xy-directions, respectively. We also implemented the inpainting mask as a brush, rendered in semi-transparent azure, which is always layered above all other colors. The second part is the parameter area (\autoref{Figure3}b), where volume fraction and load direction are entered via text. Additionally, there is a strength parameter to control the intensity of the repainting. While optional during practical use, our testing indicates the optimal value is between 0.6 and 0.8. Finally, there is the canvas area (\autoref{Figure3}c). The left canvas is used for sketch input, while the right canvas displays the generated results. Generation can be initiated by pressing the "generate" button in the bottom left corner after drawing is complete.

A typical case is illustrated in right part of \autoref{Figure2}. First, users use brush tools to draw the material region and physical constraints. Next, they can choose whether to use the mask function to generate results in a specific area; if no mask is applied, the system performs topology optimization over the entire material region by default. Users then define the volume fraction and the load direction using text inputs. Finally, they click the Generate button to run topology optimization, and can repeat the generation process for rapid iteration.

\subsection{Technical Details}

One advantage of Sketch2Topo is that it does not require retraining the underlying foundation model; instead, it only involves modifying the working pipeline of the diffusion model, and can seamlessly incorporate more powerful foundation models in the future. We adopt the open-source TopoDiff as the base model~\cite{maze2023diffusion}. Aside from user-configurable parameters, all other parameters are fixed: the filter radius is set to $r_{\mathrm{min}} = 2.0$, Poisson's ratio to $\nu = 0.3$, and the penalty factor to $p = 3.0$ and the Young's modulus is set to $E_0 = 1$ for the solid phase and $E_{\mathrm{min}} = 10^{-9}$ for the void phase.

First, to support the input of objects with specific shapes, we incorporated an image-to-image approach into the base model. The core idea is to first apply a certain level of noise to the input image and then denoise it back, without requiring any fine-tuning of the model~\cite{saharia2022palette}. Regarding the input of physical constraints, we utilize distinct colors. We employ OpenCV to identify color and coordinate information within the image and convert them into usable parameters~\cite{opencv_library}. Finally, for the inpainting functionality, the mask brush in the graphical user interface is used to define the masked regions. All generation tasks are performed exclusively within the masked areas, while the remaining regions are kept unchanged~\cite{meng2021sdedit}. Volume fraction refers to the amount of material retained relative to the initial state; for example, a value of 0.2 implies that topology optimization will remove 80\% of the material, retaining only 20\%.

\section{evaluation and APPLICATION}

\subsection{Minimum Compliance and Volume Fraction}

\begin{table*}[t]
\centering
\caption{Comparison of minimum compliance and volume fraction across different methods.}
\label{tab:compliance_vf}
\begin{tabular}{lcccc}
\toprule
 & FEA & i2i & mask 1 & mask 2 \\
\midrule
Minimum Compliance & 63.40 & 94.97 $\pm$ 30.64 & 96.49 $\pm$ 48.77 & 67.96 $\pm$ 18.71 \\
Volume fraction (\%) & 20 & 21.90 $\pm$ 0.67 & 24.76 $\pm$ 1.49 & 26.34 $\pm$ 1.55 \\
\bottomrule
\end{tabular}
\end{table*}

To preliminarily verify whether Sketch2Topo effectively contributes to mechanical performance, we conducted a brief quantitative evaluation and comparative study on minimum compliance and volume fraction. In total, four groups were evaluated: traditional finite element analysis (FEA), image-to-image generation without masking, and two different masking conditions. The boundary conditions are identical to those shown in \autoref{Figure1}: the coordinate origin is set at the lower-left corner of the design domain; a vertical downward load is applied at $(0.98, 0.96)$; and fixed supports are placed at $(0.26, 0.0)$ and $(0.0, 0.62)$. The prescribed volume fraction is set to $0.2$. The four groups correspond to \autoref{Figure1} a, c, d, and e, respectively. Apart from these settings, all other parameters are kept consistent across experiments: the filter radius is set to $r_{\mathrm{min}} = 2.0$, Poisson's ratio to $\nu = 0.3$, the penalty factor to $p = 3.0$, and the Young's modulus is set to $E_0 = 1$ for the solid phase and $E_{\mathrm{min}} = 10^{-9}$ for the void phase. For the generative AI-based methods, we perform 10 independent generations for each group and report the mean and standard deviation. The quantitative results are summarized in \autoref{tab:compliance_vf}.

Overall, the structures generated by Sketch2Topo exhibit higher minimum compliance compared to those obtained via traditional finite element analysis. However, this increase in compliance is inevitable due to the additional regional constraints introduced by hand-drawn inputs and remains within an acceptable range. Furthermore, we observe notable differences in volume fraction between the mask-based results and the traditional FEA baseline. This is primarily because the generative model tends to preserve materials that contribute less to structural stiffness; to maintain favorable structural performance, the model sacrifices strict adherence to the prescribed volume fraction in favor of minimizing compliance. This behavior reflects the higher relative weighting of minimum compliance compared to volume fraction in the model’s optimization objective.

\subsection{Iterative Design of a Chair}

\begin{figure*}
\centering
\includegraphics[width=\linewidth]{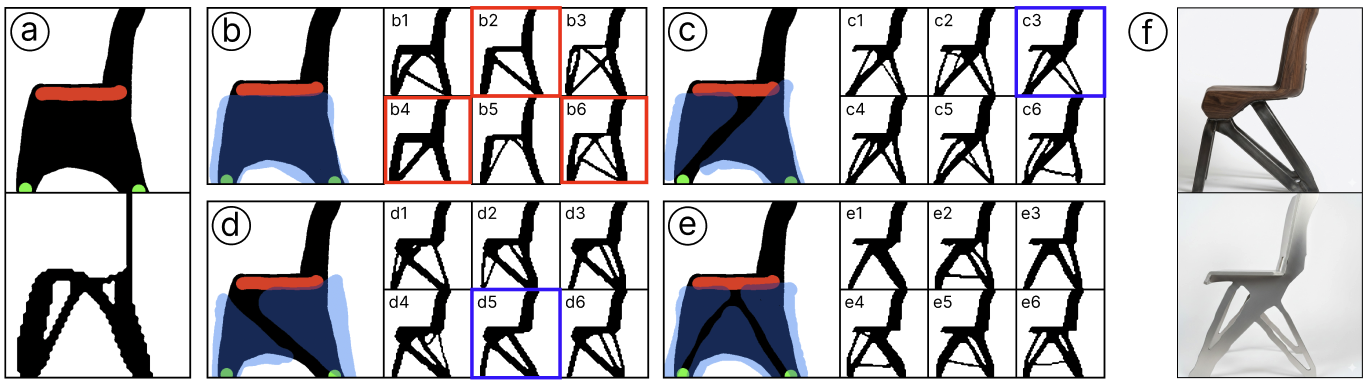}
\caption{Iterative chair design using Sketch2Topo: (a) without the mask function; (b) with the mask function for creative exploration; (c)–(e) topology optimization results under three different mask conditions; and (f) the final rendered design (rendered using Nano Banana from google).}
\label{Figure4}
\end{figure*}

In many real-world scenarios, design does not solely pursue extreme material efficiency, but instead seeks a subtle balance with aesthetic considerations. Sketch2Topo aims to satisfy performance requirements while placing greater emphasis on additional factors such as dimensions and aesthetic qualities. We therefore conducted a form-finding experiment on chair legs using Sketch2Topo. As shown in \autoref{Figure4}, the load direction was vertically downward, and the volume fraction was set to 0.15. In the first round, we drew the constraints as shown in \autoref{Figure4}a and performed 10 generations. However, we observed that the seat surface deformed during the generation process. In the second round, we used the mask function to restrict topology optimization to the chair leg region only (as shown in \autoref{Figure4}b). We performed 20 generations and present six results. Since diffusion-based models tend to produce similar outputs, we only display the most distinct results; the same selection criterion is applied hereafter. In the third step, inspired by the generated results (highlighted by the red bounding boxes in \autoref{Figure4}b), we redrew three different masks to further control the generation outcomes (\autoref{Figure4}c, d, and e). Finally, we invited 10 product designers to vote on the generated designs and selected the two chairs with the highest vote counts (\autoref{Figure4}c3 and d5). We then used Nano Banana image generation model from google to produce product-style renderings of the selected designs, as shown in \autoref{Figure4}f.

\section{CONCLUSION}
In this paper, we present Sketch2Topo, an interactive design tool that integrates hand-drawn sketch input into a diffusion-based topology optimization workflow, aiming to address the lack of user control in existing data-driven approaches. By leveraging diffusion models with inpainting and sketch-conditioned control capabilities, our system enables users to intuitively define physical constraints and geometric features through direct manipulation. We further illustrate the potential value of Sketch2Topo through a brief quantitative study and an application example. Our approach also has several limitations. Currently, sketch-based inputs primarily serve as guidance for topology optimization, and designs intended for fabrication still require manual reconstruction. In addition, the system currently operates mainly in a two-dimensional domain, which limits its direct applicability to complex volumetric components. In future work, we plan to train new models using expanded datasets and explore integrating the system into 3D design software to further enhance the interactive experience.

\begin{acks}
This work was supported by Japan Science and Technology Agency (JST) as part of Adopting Sustainable Partnerships for Innovative Research Ecosystem (ASPIRE), Grant Number JPMJAP2401.
\end{acks}

\balance
\bibliographystyle{ACM-Reference-Format}
\bibliography{references}

\end{document}